\def\kpc{ {\rm kpc} }
\def\pc{ {\rm pc} }
\def\pcm3{ {\rm pc}^{-3} }

\def\fha{ f_{\rm h} }
\def\kms{ {\rm km} {\rm s}^{-1}  }
\def\msun{ {\rm M}_{\sun} }
\def\Rsun{ {\rm R}_{\sun} }

\def\Rc{R_{\rm c}}
\def\Rs{R_{\rm s}}
\def\ms{m_{\rm s}}
\def\sech{{\rm sech}}
\def\Rp{R_{\rm p}}
\def\Rcal{{\cal R}}
\def\Pcal{{\cal P}}
\def\vproj{v_{\rm p}}
\def\aproj{a_{\rm p}}
\def\be{ \begin{equation} }
\def\ee{ \end{equation} }
\def\half{{\frac{1}{2}}}

\def\spose#1{\hbox to 0pt{#1\hss}}
\def\lta{\mathrel{\spose{\lower 3pt\hbox{$\sim$}}
    \raise 2.0pt\hbox{$<$}}}
\def\gta{\mathrel{\spose{\lower 3pt\hbox{$\sim$}}
    \raise 2.0pt\hbox{$>$}}}

\documentstyle[12pt,aasms4,epsf,rotate]{article}

\begin{document}

\title{A Caustic View of Halo Microlensing}

\author{Eamonn J. Kerins}
\and
\author{N. Wyn Evans}
\affil{Theoretical Physics, Department of Physics, 1 Keble Rd, Oxford,
OX1 3NP, UK}

\begin{abstract} 
The only microlensing events towards the Magellanic Clouds for which
the location of the lens is strongly constrained are the two binary
caustic crossing events. In at least one and possibly both cases, the
lens lies at, or close to, the Magellanic Clouds themselves.  On the
face of it, this seems an improbable occurrence if the Galactic dark
halo provides the bulk of the lensing population, as suggested by
standard analyses of the MACHO dataset towards the Large Magellanic
Cloud (LMC). We use a binomial statistic to assess the prior
probability of observing $M$ non-halo binary caustic events given a
total sample of $N$ caustic binaries. We generalize for the case of
multi-component Galactic and Magellanic Cloud models the Bayesian
likelihood method for determining the lens mass and halo fraction from
the observed timescales. We introduce a new statistic, the ``outcome
discriminator'', which measures the consistency between the binary
caustic data, prior expectation, and the MACHO two-year LMC dataset as
a whole.

If the Magellanic Clouds are not embedded in their own dark halos of
MACHOs, then the discovery of two non-halo caustic binary events out
of two ($M = N = 2$) is inconsistent with expectation given the MACHO
dataset. Galactic models in which $M = 1$ is the likeliest outcome are
also inconsistent with the data, though models in which $M = 1$ has a
reasonable prior probability are not. We consider the possibilities
that the Magellanic Clouds are embedded in dark haloes of their own,
or that the Galactic halo is intrinsically deficient in the binary
systems which produce caustic crossing events. Either of these
possibilities provide greater compatibility between observation and
prior expectation, though the idea of Magellanic haloes is perhaps the
more natural of the two and has support from kinematical studies.
\end{abstract}


\section{INTRODUCTION}

The main motivation driving the microlensing experiments towards the
Magellanic Clouds has been the identification of the dark matter
content of our Galactic halo.  Although there is now no doubt that
microlensing events are being detected, their origin is still highly
controversial. The sheer number of events seems to argue in favor of
the lenses lying in the Galactic halo, but the timescales do not lend
themselves to any easy interpretation in terms of known astrophysical
objects. This has led a number of investigators to suggest other
possible locations for the lenses, including the Magellanic Clouds
themselves (Sahu 1994), intervening stellar populations (Zaritsky \&
Lin 1997), tidal debris (Zhao 1998) or even a warped and flaring Milky
Way disk (Evans et al. 1998).

Exotic microlensing, such as parallax and binary events, allow us to
break some of the degeneracies in the mass, distance and transverse
velocity of the lens (Mao \& Paczy\'nski 1991; Gould 1992). In
particular, binary caustic crossing events possess a striking
lightcurve, enabling unambiguous identification. Even better, the
stellar radius crossing time $t_\star$ enables the projected
transverse velocity $\vproj = \Rs/ t_\star$ at the source to be
measured, provided the source radius $\Rs$ is known.  The current
follow-up searches of microlensing alerts allow for dense sampling of
the lightcurves and permit resolution of the caustic crossing (e.g.,
Alcock et al. 1997a; Albrow et al. 1998). Thus far, two such binary
caustic crossing events have been detected. The first, MACHO~LMC-9,
had $t_\star = 0.65$ days and was found by Bennett et al. (1996) for a
source star in the Large Magellanic Cloud (LMC). The second and most
recent, 98-SMC-1, occurred towards the Small Magellanic Cloud (SMC) and was
intensively monitored by three groups (Afonso et al. 1998; Albrow et
al 1999; Alcock et al. 1999), who found $t_\star = 0.12$ days.

Section 2 discusses the likely location of these events. At least one,
and possibly both, reside outside of the Galactic halo. Given that
halo lenses are expected to outnumber greatly other lens populations,
this appears to be a remarkable result. This is confirmed in Section
3, which introduces a binary caustic crossing statistic measuring the
prior expectation that $M$ out of a sample of $N$ binary caustic
crossing events lie outside of the Galactic halo. Section 4 presents
minimal models for the Magellanic Clouds as a prelude to assessing the
consistency between prior expectation and observation in Section 5.
Here, a new statistic -- the ``outcome estimator'' -- is introduced
and used to show that there are no minimal models that are consistent
with the observed timescales and in which $M = 1$ or $2$ is the
likeliest outcome, although there are compatible minimal models in
which $M = 1$ has a reasonable prior probability. Two alternative
possibilities -- namely that the Magellanic Clouds are swathed in
their own dark halos or that the Galactic halo is under-endowed with
binaries -- are considered in Sections 6 and 7. Finally, we consider
the implications of the discovery of further binary caustic crossing
events in Section 8.

\section{LOCATION OF THE BINARY CAUSTIC EVENTS}

For the SMC event, the observed flux and inferred effective
temperature of the source suggest that the stellar radius $\Rs = 1.2 \,
\Rsun$. Thus, $\vproj  = \Rs / t_\star \simeq 80 \,\kms$ (Alcock et
al. 1999). Recalling that the halo's optical depth peaks at
heliocentric distances of $\sim 10\,\kpc$, it is easy to calculate
that typical halo lenses have $\vproj \sim 1000 \, \kms$. The low
$\vproj$ for the SMC event seems to make it certain that the lens
responsible does not reside in the Galactic halo but instead belongs
to the SMC itself.

The source for the LMC event appears to be a A7-8 main sequence star
($\Rs \simeq 1.5 \,\Rsun$), in which case $\vproj \simeq 20 \, \kms$
(Bennett et al. 1996). Such a value for $\vproj$ seems unusually low
even for LMC lenses. This led Bennett et al. (1996) to suggest that
the source itself may be binary, comprising roughly equal luminosity
companions with a projected separation $\aproj$ much less than the
projected Einstein radius of the lens $\Rp$. In such an instance, the
crossing time would be given by $t_* = \aproj/\vproj$.  To see how
this might affect the inferred location of the binary lens, let us
first note from Kepler's third law that
   \be
      \aproj \leq a = 2.5 \, \mbox{AU} \, \left( \frac{P}{10^3 ~\mbox{d}}
      \right)^{2/3} \left( \frac{\ms}{\msun} \right)^{1/3} \label{kep}
   \ee
for a binary system of semi-major axis $a$ and orbital period $P$ comprising
stars of equal mass $\ms$. The Einstein radius of the lens projected onto the
source plane is given by
   \be
      \Rp = 40 \, \mbox{AU} \, \left( \frac{m}{\msun} \right)^{1/2} \left(
        \frac{D}{200~\mbox{kpc}} \right)^{1/2}, \label{rproj}
   \ee
where $m$ is the lens mass and $D = d_{\rm os} d_{\rm ls}/d_{\rm ol}$,
with $d_{\rm os}$, $d_{\rm ls}$ and $d_{\rm ol}$ being the
observer--source, lens--source and observer--lens distances,
respectively.  Fitting to data tabulated in Zombeck (1990),
main-sequence sources between $0.4 - 4~\msun$ can be characterized by
the following mass--radius relation:
   \be
      \Rs \simeq \left( \frac{\ms}{\msun} \right)^{4/5} \Rsun. \label{mr} 
   \ee
In order that both binary stars be lensed, we require $\aproj \ll
\Rp$.  Hence, we set $\aproj < f \Rp$ ($f \ll
1$). Equations~(\ref{kep}), (\ref{rproj}) and (\ref{mr}) therefore
give the binary semi-major axis in units of stellar radius as
   \be
      \frac{\aproj}{\Rs} \la 7 \times 10^3 \, f \left( \frac{m}{\msun}
      \right)^{1/2} \left( \frac{D}{200~\mbox{kpc}} \right)^{1/2} \left(
      \frac{\ms}{\msun} \right)^{-4/5}. \label{ap}
   \ee
Adopting $f = 0.1$, $m = 0.5~\msun$ and $\ms = 2~\msun$ gives $\aproj
/ \Rs \la 270$. So, whilst taking $\vproj = \Rs / t_\star$ gives a
rather small value of $20 \, \kms$, if the source is binary one should
use $\aproj$ in place of $\Rs$, which would make $\vproj$ up to 2
orders of magnitude larger. This would make it easily compatible with
the $\vproj \sim 1000 \, \kms$ predicted for halo lenses. Thus, whether
the LMC caustic crossing event resides in the halo or not depends
crucially on whether the source is binary or not.

What is the likelihood that the source is a binary system with the
correct characteristics? The binary fraction is a function of the
orbital period $P$. The constraint $\aproj < f \Rp$ limits $P$ to
   \be
      P < 6.6 \times 10^4 \, {\rm d} \, f^{3/2} \left( \frac{m}{\msun}
      \right)^{3/4} \left( \frac{D}{200~\mbox{kpc}} \right)^{3/4} \left(
      \frac{\ms}{\msun} \right)^{-1/2}. \label{plim}
   \ee
For the values of $f$, $m$ and $\ms$ assumed above,
equation~(\ref{plim}) gives $P \leq 900$~d. From figure~5 of
Abt~(1983), A4--F2 stellar systems with periods below 1000~d comprise
on average 1.27 members (or 0.27 companions per primary). If this
statistic is assumed to be due entirely to single and binary systems,
then it implies a binary fraction of $27\%$. Only $39 \%$ of A type
binaries comprise roughly equal mass stars (Duquennoy \& Mayor 1991;
Griest \& Hu 1992), giving just $10 \%$ as the fraction of sources
which are binaries with the correct configuration.  However, some of
the systems tabulated by Abt (1983) comprise more than two members, so
in fact this figure is still an upper limit. Assuming these numbers
apply to the LMC, then there is no more than a one in ten chance that
the crossing time is indicative of the separation between binary
companions rather than the size of the stellar disk.

Bennett et al. (1996) put forward the intriguing suggestion of a
binary source because they felt that the value of $\vproj \simeq 20
\, \kms$ seemed rather low. If the one-dimensional velocity dispersion
of the LMC disk is $\sim 25\,\kms$ then, as we show in section~4, the
most likely value for $\vproj$ is around $55 \, \kms$; a velocity as
low as $20 \, \kms$ is only one-tenth as likely.  But, this conclusion
does depend sensitively on the assumed velocity dispersion. If the
velocity dispersion is $\sim 15\,\kms$, then the relative probability
is more appreciable at $\sim 25 \%$ of the peak value. The velocity
dispersions of the young, intermediate-age and old stellar populations
in such a complex system as the LMC are likely to differ, so it is
difficult to judge which value is the most appropriate.  We believe
that a definite conclusion as to the location of the lens of the LMC
binary caustic crossing event cannot be reached with the available
evidence. The question could be unambiguously settled by spectroscopy
at multiple epochs to see if the source is a radial-velocity variable
and hence a binary.  Depending on the interpretation of the LMC event,
the present observations may be represented by $M = N = 2$ or by $M=1,
N =2$. Both possibilities are given equal consideration in what
follows.

\section{BINARIES AND BINOMIALS}

Consider a dark Milky Way halo of which some fraction comprises
lenses. Of course, there are other populations of lenses along the
line of sight to the Clouds -- in particular, the Magellanic Clouds
themselves and the Milky Way disk. These we collectively refer to as
non-halo lenses.  Let $R$ (``the rate factor'') be the factor by which
the rate of microlensing events caused by halo lenses exceeds the rate
of non-halo lenses.  In each population, some of the lenses will be
binaries, some will not. Only a subset of binaries with particular
configurations will give rise to the rather special caustic crossing
events.  The number of binary caustic crossing events as compared to
the total number of all events in each population is unknown, either
for the Galactic halo or for non-halo populations.  The simplest
assumption to make is that this fraction is the same in all
components.  This may not be the case. First, the fraction of binary
stars may vary between stellar populations, second, the distributions
of semi-major axes and mass ratios may change according to
environment, and third, the caustic-crossing cross section varies with
location. Another possibility is a selection bias against events with
short intervals between caustic crossings, the majority of which are
expected to be caused by halo lenses (Honma 1999).  To allow for such
effects, let us define $B$ (``the binary bias'') as the ratio of halo
to non-halo caustic crossing binary fractions.  In other words, if
$B=1$, then the fraction of all events that are caustic crossing is
the same for all stellar components.  If $B=\half$, then the non-halo
caustic crossing binary fraction is twice as large as that of the
halo.

The binomial prior probability of observing $M$ non-halo
caustic crossing binaries given a total sample of $N$ such events is
then
\be
\Pcal(M|N) = \frac{N!}{M!(N-M)!} \left( \frac{1}{B R + 1}
\right)^M \left( \frac{B R}{B R + 1} \right)^{N-M}. \label{e2}
\ee
The rate factor $R$ contains all the Galactic modeling information,
namely the density contribution, velocity distributions and mass
functions of the lenses, and also the experimental efficiencies.  The
two caustic crossing events appeared in different locations (the first
towards the LMC, the second towards the SMC) which complicates the
computation of $R$.  The experimental efficiencies for the SMC
observations are not yet available, so it is not possible to assess
$R$ along that line of sight.  We assume $R$ to be the same along both
the LMC and SMC lines of sight.  Even if this turns out not to be a
good assumption, our results are conservative as long as $R$ towards
the SMC is at least as large as $R$ towards the LMC. Since the LMC is
substantially more massive than the SMC, this is a reasonable
assumption provided that the SMC is not too elongated along the line
of sight (see Palanque-Delabrouille et al. 1998). So, if anything, we
are overestimating the importance of the Clouds (though see Sahu \&
Sahu (1998) for a contrary viewpoint).

Let us first give an intuitive feel for the meaning of
equation~(\ref{e2}).  Somewhat anticipating the results of the next
Section, let us take $R = 5.5$. Displayed in figure~1 is the resulting
binomial probability $\Pcal(M|N)$ for $1 \leq N \leq 4$ assuming a
binary bias $B=1$ (i.e. no bias).  We see in figure~1 that for the
above parameters $\Pcal(1|2)$ is $26 \%$, whilst $\Pcal(2|2)$ is very
low at $2.4\%$. If the source for the LMC event is indeed binary, thus
allowing this event to lie in the halo, then there is nothing
remarkable about the present observational status. But, if the source
is not binary, then both events are almost certain to be of non-halo
origin. In this case, the small prior probability $\Pcal(2|2)$ does
suggest that something is amiss.  If a third binary caustic crossing
event were to be discovered and found to be non-halo, then the prior
probabilities are $\Pcal(2|3)$ at $6\%$ and $\Pcal(3|3)$ at just
$0.4\%$. Interestingly, only when $N > 4$ does the probability of $M >
0$ exceed that for $M = 0$.

For larger samples (i.e. larger $N$), the probability of any specific
outcome $\Pcal(M|N)$ inevitably decreases because of the increase in
the range of possibilities. For this reason, it is preferable to
normalize probabilities with respect to the most likely outcome
$\Pcal_{\rm max}(N)$ for a given sample $N$.  In the case of $N = 2$, the
most likely outcome is $M = 0$ with $\Pcal_{\rm max} (2) = \Pcal(0|2)
= 0.71$ for our assumed $R$. Let us therefore define $\Rcal(M|N) = \Pcal(M|N) /
\Pcal_{\rm max}(N)$ as the prior relative probability. Thus for $N = 2$,
$\Pcal_{\rm max}(2) = \Pcal(0|2)$, so we have $\Rcal(0|2) = 1$,
$\Rcal(1|2) = 0.37$ and $\Rcal(2|2) = 0.033$. Even relative to other
possible outcomes, two non-halo caustic crossing events from a sample
of two is clearly unexpected. 

Whilst interesting, these figures are in no sense definitive since
they depend on a number of uncertain modeling parameters, which
determine the rate factor $R$. The models themselves are of course
constrained by the entire microlensing dataset. We therefore undertake
a more critical examination by first introducing Galactic and LMC
models and then more elaborate statistical techniques in the following
two Sections.

\section{MINIMAL MAGELLANIC MODELS}

The simplest model for the LMC is a bare disk with a central column
density of $363\,\msun\pc^{-2}$ and an inclination angle of $27^\circ$
(e.g., Alcock et al. 1996; Westerlund 1997). The mass density of the
LMC disk is taken as (c.f., Evans 1996)
\begin{equation}
\rho(R,z) = 0.54 \exp\left(- 0.625 \sqrt{R^2 + 28.6 z^2} \right)
\msun\pc^{-3},
\label{minlmcmodel}
\end{equation}
where ($R,z$) are the cylindrical coordinates, in kiloparsecs,
centered on the LMC. The scale length of the LMC disk is $1.6\,\kpc$
and the scale height is $300\,\pc$ (Bessell, Freeman \& Wood 1986).
The velocity distribution is isotropic and Gaussian with a
one-dimensional velocity dispersion of $25\,\kms$ about an asymptotic
circular velocity of $80\,\kms$ (e.g., Schommer et al. 1992;
Westerlund 1997).  The Galactic halo is represented as an isothermal
sphere of core radius $\Rc =5\,\kpc$ and an asymptotic circular speed
$v_0$ of $220\,\kms$ (c.f., Griest 1991):
\begin{equation}
\rho(r) = {v_0^2 \over 4\pi G (\Rc^2 + r^2)},
\label{kimshalo}
\end{equation}
where $r$ is a spherical polar coordinate measured from the Galactic
Center.  The velocity distribution is taken to be isotropic and
Gaussian with a dispersion of $155\,\kms$. The Galactic disk is
modeled by 
\begin{equation}
\rho(R,z) = 0.62 \exp\left(-\frac{R}{3\,\mbox{kpc}}\right)
\sech^2\left(\frac{z}{0.3\,\mbox{kpc}}\right)\msun\pc^{-3},
\label{mwdisk}
\end{equation}
where ($R,z$) are cylindrical polar coordinates about the Galactic
Center. The column density of the thin disk at the sun is
$26\,\msun\pc^{-2}$, as suggested by Gould, Bahcall \& Flynn
(1997). The velocities are distributed about the circular speed of
$220\,\kms$ in Gaussian manner with a dispersion of $30\,\kms$.  The
motion of the line of the sight is taken into account using the proper
motion measurement of Jones, Klemola \& Lin (1994). The LMC and Milky
Way disk lenses are taken to have a fixed mass of $0.35\,\msun$, since
this represents the average microlensing mass for a hydrogen-burning
population with a Scalo mass function. The Milky Way halo lenses are
characterized by a discrete lens mass $m$.  With these model
assumptions in hand, it is now straightforward to calculate the rate
of microlensing.

Figure~2a plots the distribution of projected velocities $P(\vproj)$
for lenses in the Milky Way disk and halo, as well as the LMC disk.
(Somewhat anticipating the work in Section 6, the results for an LMC
dark halo are also included). Each of the distributions is normalized
to have a maximum value of unity. As Bennett et al. (1996) articulate,
this is potentially a powerful way to separate the binary caustic
crossing events by the location of the lenses. For example, lenses in
the Milky Way disk typically lie within heliocentric distances of a
kiloparsec, so the projection factor is $\gta 50$ and the projected
velocities of the lenses $\gta 2000\,\kms$.  The projection factor of
lenses in the LMC disk slightly exceeds unity, but the typical
projected velocity is $\sim 55\,\kms$, a good deal higher than the
two-dimensional velocity dispersion. This is because the random
velocities of both the lenses and sources contribute and because the
event rate is determined by the flux which is velocity weighted.
Figure~2b assesses how likely it is that an LMC lens has $\vproj
\simeq 20 \,\kms$.  Here, the relative probability density $P(\vproj =
20)$ is plotted against the velocity dispersion $\sigma$ of the LMC
disk. The unbroken line refers to a LMC disk with the standard scale
height of $300\,\pc$, the broken line refers to a super-thin disk with
half the standard thickness. Whether such a low $\vproj$ as $20\,
\kms$ is likely or unlikely depends sensitively on the assumed
velocity dispersion and scale height.

\section{THE OUTCOME ESTIMATOR}

There are only two free parameters that remain in our model. The first
is the lens mass $m$ and the second is the fractional contribution to
the halo provided by the lenses, which we denote as $\fha$.  For a
specified Galactic model, both $\fha$ and $m$ are constrained by
microlensing observations. As $\Rcal(M|N)$ depends upon these
parameters, it too is therefore restricted by the data.  Alcock et
al. (1997b) use a Bayesian likelihood statistic to constrain both
$\fha$ and $m$, assuming the observed events all reside in the
Galactic halo. We drop this assumption by generalizing their statistic
as follows
\be 
L(\fha,m|t_i) =
\exp(-\fha N_{\rm h} - N_{\rm d}) \prod_{i = 1}^{N_{\rm obs}} E {\cal
E}(t_i) \left( \fha \frac{{\rm d}\Gamma_{\rm h}}{{\rm d}t_i} +
\frac{{\rm d}\Gamma_{\rm d}}{{\rm d}t_i} \right), 
\label{like}
\ee 
where $N_{\rm obs} = 8$ is the observed number of events, $E = 1.82
\times 10^7$ star-years is the effective exposure for the Alcock et
al. 2-year dataset, ${\cal E}(t_i)$ is the detection efficiency for
events of duration $t_i$, $N_{\rm h}$ and $\Gamma_{\rm h}$ are
respectively the expected number of events and the event rate for a
full halo ($\fha = 1$), and $N_{\rm d}$ and $\Gamma_{\rm d}$ are the
analogous quantities for the combined Milky Way and LMC disk
components. Figure~3a plots the resulting likelihood contours (solid
lines) in the ($\fha,m$) plane assuming a uniform prior in $\fha$ and
logarithmic prior in $m$. The effect of including the non-halo
components is that our two dimensional maximum likelihood solution
($\fha = 0.49, m = 0.50\, \msun$, as indicated by the cross) implies a
baryonic halo mass inside $50\,\kpc$ of $2.3 \times 10^{11}\,\msun$;
this is $80\%$ of the mass derived in the Alcock et al. (1997b)
analysis for the same events. \footnote{Alcock et al. (1997b) do
derive a similar halo mass using a fixed six event ``halo''
subsample. However, the likelihood expression in equation~(\ref{like})
has the advantage that it is robust to variations in the assumed
model, such as a more massive disk embedded in a lighter halo.}
Marginalizing over $\fha$ gives an allowed range in $m$ of $0.34 -
1.03\,\msun$ ($0.19 - 1.82\,\msun$) at the $68 \%$ ($95 \%$)
confidence level. For $\fha$, marginalizing over $m$ gives an allowed
range of $0.26 - 0.79$ ($0.12 - 1.26$).

The shaded regions in figure~3a show the most likely number of
non-halo caustic crossing events $M$ given $N = 2$ observed binary
caustic events. By taking $N = 2$, we are including the SMC binary
event although the likelihood contours do not incorporate this
event. This is justifiable so long as the LMC dataset is truly
representative of the halo lens population as a whole. (As we stressed
earlier, we are also assuming that the rate factor $R$ is similar
along both the LMC and SMC lines of sight). For $(\fha,m)$ values
falling in the darker shaded region to the upper left of the plot,
$\Rcal(2|2) = 1$. Thus, the prior expectation is $M = 2$ non-halo
events (i.e. none of the events reside in the halo) for models in this
region. In the lighter shaded region labeled ``$M = 1$'', $\Rcal(1|2)
= 1$, whilst in the lightest shaded region $\Rcal(0|2) = 1$.

At this point it is worth stressing the difference between the model
likelihood $L$ and the binary prior relative probability
$\Rcal(M|N)$. A value of $\Rcal (M|N)$ approaching unity means that
$M$ non-halo caustic crossing binaries are consistent with expectation
given that $N$ caustic crossing events have been observed. A large $L$
means that the Galactic model (characterized by $m$ and $\fha$) is
consistent with the timescales $t_i$ of all the LMC microlensing
candidates. $\Rcal(M|N)$ is a {\em model dependent}\/ quantity, since
$\fha$ and $m$ are required to calculate $\Pcal (M|N)$ in
equation~(\ref{e2}). However, $\Rcal (M|N)$ {\em does not}\/ provide a
measure of the likelihood of the underlying model $(\fha,m)$ itself;
it is meaningless to compare values of $\Rcal(M|N)$ between models
with different $(\fha,m)$ without taking account of the model
likelihood $L$ in each case. The likelihood $L$ can be used to assess,
in a {\em model independent}\/ manner, the relative likelihood of the
hypotheses that we inhabit a galaxy in which $M = 0$, 1 or 2 is the
expected outcome for the case $N = 2$. Mathematically speaking, we can
use $L$ to test the hypotheses $\Rcal(0|2) = 1$, $\Rcal(1|2) = 1$ or
$\Rcal(2|2) = 1$. It is evident from the solid contours that most of
the likelihood $L$ is contained within the region where $\Rcal(0|2) =
1$. Integrating the likelihood in each region we find that, given the
data, $M = 0$ is 40 times more likely to be the expected outcome than
$M = 1$, and 1700 times more likely than $M = 2$.

Whilst we may expect the present observational situation to have a
significant prior probability, it would be unreasonable of us to
demand that it be the {\em most}\/ probable. Since the current caustic
binary dataset indicates $N = 2$ and $M = 1$ or 2, we wish to see
whether there are Galactic models that predict significant values for
either $\Rcal (1|2)$ or $\Rcal (2|2)$ (or both), and which are
consistent with the entire dataset (i.e. have a large likelihood
$L$). We therefore demand that $\Rcal(M|N)$ be at least as large as
some threshold $\Rcal_*$ where, say, $\Rcal_* = 0.1$. In other words,
we require the observations to have a prior probability that is at
least one-tenth as large as that of the expected outcome. We can again
test this hypothesis for each $M$ by integrating the likelihood $L$
over the regions where $\Rcal(M|N) \geq \Rcal_*$. We therefore wish to
compute the ``outcome discriminator''
\be
      D(\Rcal \geq \Rcal_*|t_i) = \frac{\int \int L(\fha,m|t_i) \,
      \Theta(\Rcal - \Rcal_*) \, {\rm d}\ln m \, {\rm d}\fha}{
      \int \int L(\fha,m|t_i) \, {\rm d}\ln m \, {\rm d}\fha}, \label{hyp}
\ee
where $\Theta$ is the Heaviside step function and uniform priors in
$\fha$ and $\ln m$ have been assumed. The outcome discriminator is the
probability that the condition $\Rcal \ge \Rcal_*$ holds given the
dataset $t_i$. Note that, in principle, the cleanest way to interpret
the data would be to use a maximum likelihood statistic in $t$ for
ordinary events and ($t, \vproj$) for the binary caustic crossing
events. This is not feasible as we would then need to know the
theoretical rate and the efficiencies for caustic crossing events
towards the Magellanic Clouds.

Figure~3b shows plots of the outcome discriminator $D(\Rcal \geq
\Rcal_*|t_i)$ against threshold $\Rcal_*$ for the case $N=2$.  As
$\Rcal_* \rightarrow 0$, the outcome discriminator tends to unity
because the fraction of the likelihood plane satisfying the threshold
$\Rcal_*$ tends to unity. As $\Rcal_*$ increases, the outcome
discriminator decreases until, at $\Rcal_* =1$, it attains the limit
corresponding to the shaded regions in figure~3a. At first glance the
curve for $M=0$ (thick solid line) appears to be insensitive to
$\Rcal_*$. Looking back at figure~3a we see that, even when $\Rcal_* =
1$, the overwhelming bulk of the likelihood is contained within the
$M=0$ region. Therefore, decreasing the threshold $\Rcal_*$ does not
significantly increase the value of the outcome discriminator.  By
contrast, when $\Rcal_* =1$, the $M=1$ and $2$ regions cover very
little of the likelihood plane. So, in these cases decreasing the
threshold $\Rcal_*$ results in a significant increase in the outcome
discriminator. Superimposed on the figure are two cuts, depicted by
the hatched regions. The vertical region denotes a minimum acceptable
threshold $\Rcal_* = 0.1$. Outcomes below this threshold are taken to
have unacceptably low prior probability.  The horizontal hatched
region shows where the condition $\Rcal \ge \Rcal_*$ has less than $10
\%$ chance of being satisfied, given the two-year LMC dataset. It is
interesting that with these adopted thresholds, $M=2$ is almost
completely excluded: there is no more than a $10\%$ chance that we
inhabit a galaxy where $\Rcal(2|2)$ exceeds 0.1. For $M=1$, models
with large prior relative probability ($\Rcal \ge 0.65$) are excluded
by the data. This however still leaves viable models in the unhatched
region.

Clearly, these figures do not favor the interpretation of both caustic
crossing events residing outside the Galactic halo. Therefore, if it
really is the case that $M=2$, something must be amiss in our model
assumptions. What are our options if this does indeed turn out to be
the case?

\section{MAXIMAL MAGELLANIC MODELS}

If the source stars of the two caustic crossing events really are
single, implying that both the lenses are indeed of non-halo origin,
then the preceding analysis leads one to think that the Galactic model
itself may be wrong. Perhaps the halo is too massive, whilst the LMC
disk and Milky Way stellar components are not substantial
enough. However, this is not an option because virial arguments
preclude the LMC disk from accounting for most of the microlensing
events (Gould 1995), whilst the projected velocities of the caustic
crossing events are too low to be explained by a Milky Way disk,
however massive.  Alternatively, there might be an additional dark
component which contributes significantly to the microlensing
statistics, thereby implying a somewhat reduced Milky Way halo
microlensing budget. Natural candidates for this are halos of dark
matter enveloping the LMC and SMC themselves -- for which there is
supporting kinematical evidence (Schommer et al. 1992).  Let us
therefore immerse the LMC disk in a dark halo using
equation~(\ref{kimshalo}) with a core radius $\Rc$ of $1\,\kpc$, an
asymptotic circular speed of $80\,\kms$ and velocity dispersion of
$55\,\kms$. The LMC halo extends to a radius of about $10\,\kpc$ and
-- like the Milky Way halo -- is assumed to have a fraction $\fha$ of
lenses of mass $m$. Figure~2 shows the expected distribution of
$\vproj$ for this model, whilst Table 1 summarizes its microlensing
properties.  

The maximum likelihood solution for the combined Milky Way and LMC
disk/halo models can be readily evaluated by augmenting the $N_{\rm
h}$ and $\Gamma_{\rm h}$ terms in equation~(\ref{like}) with the LMC
halo contribution.  Figure~4a shows the resulting likelihood contours
in the ($\fha, m$) plane. There is now an additional source of lenses,
so the likelihood contours are shifted towards lower values of $\fha$.
The two dimensional maximum likelihood solution shown by the cross is
($\fha = 0.37, m = 0.43\,\msun$) corresponding to a microlensing halo
mass inside $50\,\kpc$ of $1.7\times 10^{11}\,\msun$.  The inclusion
of an LMC halo lowers the lens mass $m$ by $18 \%$ and the halo mass
by $24\%$ relative to the minimal Magellanic model.  The marginalized
distributions provide the following $68\%$ ($95\%$) limits: $\fha$
lies between $0.20 - 0.61$ ($0.09 - 0.99$) and $m$ lies between $0.29
- 0.92\, \msun$ ($0.16 - 1.79\, \msun$). As in figure~3a, we have
shaded the areas of the likelihood plane where $M=0,1$ or 2 is the
most probable prior for $N = 2$. The most significant change is that
the $M=1$ region now occupies a substantial share of the likelihood.
Integrating the likelihood in each region we find that, given the
data, $M = 0$ is twice as likely to be the expected outcome than $M =
1$, and 600 times more likely than $M = 2$.

Figure~4b shows the outcome discriminator, defined in
equation~(\ref{hyp}), as a function of threshold $\Rcal_*$. We see
that $M=0$ and $M=1$ are easily compatible with our adopted
cuts. Either of the scenarios are consistent with prior expectation
and the observed timescale distribution, though the caustic crossing
times of the binary events themselves rule out $M=0$ (see section 2).
For thresholds satisfying $0.4 \lta \Rcal_* \lta 0.7$, the $M=1$ curve
lies above the $M=0$ line. This is because the region of the
$(\fha,m)$ plane satisfying these thresholds contains a larger share
of the total likelihood for $M=1$ than for $M=0$. However, if we
tighten the threshold (i.e., increase $\Rcal_*$), then the likelihood
for $M=1$ shrinks faster than for $M=0$. A significant part of the
$M=2$ curve now lies in the permitted region in figure~4b, though the
observed timescale distribution strongly disfavors models in which
$M=2$ is {\it a priori} the most probable outcome. Such models
represent only $0.1 \%$ of the total likelihood. However, there are
still models favored by the 2-year LMC dataset in which, although it
does not have the largest prior expectation for $N = 2$ binary events,
$M=2$ is not highly improbable.

In summary, introducing an LMC halo allows models with $M=1$ as the
most probable prior for $N = 2$ to be compatible with the LMC
dataset. Even models in which $M=2$ has a reasonable prior probability
are acceptable.

\section{BINARY BIAS}

Another way of producing a relatively high proportion of non-halo
binary events is to assume that the halo is intrinsically deficient in
binary systems of the configuration required for caustic crossing to
be observed. This may be the case if the halo simply has a lower
percentage of binary stars, or if the distribution of binary
separations or mass ratios is less than optimal. In
equation~(\ref{e2}), we allowed for this possibility by introducing the
binary bias parameter $B$ as the ratio of halo to non-halo caustic
crossing binary fractions. Thus, for models in which the halo is
deficient in producing caustic crossing events $B < 1$. Let us assume
that, for whatever reason, the non-halo components have a caustic
crossing binary fraction that is twice as large as that in the halo,
so that $B = \half$. What difference does this make to our results?

Figure~5a shows the regions of the $(\fha,m)$ plane in which $M = 0$,
1 and 2 events are expected for the case $B = \half$ and $N = 2$. As
for figure~3a, the likelihood contours for the 2-year Alcock et
al. (1997b) LMC dataset are shown assuming a minimal Magellanic
model. Comparison of figures~3a and 5a show that a large binary bias
can -- to an extent -- help to explain current results. Whilst $M = 0$
still encompasses most of the likelihood with $B = \half$, we are now
only four times more likely to be living in a galaxy in which $M = 0$
is expected than one in which $M = 1$ is the most probable prior,
though we are 270 times more likely to be living in such a galaxy than
one in which $M = 2$ is expected. This compares to the case of $B = 1$
in figure~3a in which we were 40 times more likely to be living in a
galaxy with $M = 0$. Looking at the outcome discriminator in figure~5b
we see that both $M = 0$ and $M = 1$ are quite compatible with prior
expectation and the data. Models in which $M = 2$ has a prior relative
probability up to $\Rcal \simeq 0.3$ are reasonably consistent with
the dataset, though this scenario is clearly not strongly favored by
the data.

\section{THE FUTURE}

Having assessed the significance of the current sample of binary
caustic crossing events for the structure of the Galaxy and the
Magellanic Clouds, we now turn our attention to future
observations. How do our results change in the event of a third binary
caustic lens system being observed?

In the case where the LMC does not have a halo of its own, figure~6a
shows the regions of the $(\fha,m)$ plane in which the prior
expectation is $M = 0$ (lightest shaded region), 1, 2 and 3 (darkest
shaded region) non-halo events for $N = 3$ binary events. Superimposed
are the likelihood contours shown in figure~3a derived from the Alcock
et al. (1997b) two-year LMC dataset.  In reality, even if future
datasets concord with the present distribution, the resulting
likelihood contours will be somewhat more concentrated about the
maximum likelihood solution than indicated in figure~6a.  Assuming for
the present purposes that the existing likelihood contours still
apply, even for a total sample of $N = 3$ binary events, the region in
which $M = 0$ is expected still occupies the largest share of the
likelihood. According to the two-year data, we are ten times more
likely to inhabit a galaxy in which $M = 0$ is the most probable prior
than one in which $M = 1$ is expected, roughly 370 times more likely
than one in which $M = 2$ is expected, and 3000 times more likely than
one in which $M = 3$ is expected.  Unless future microlensing events
produce a shift in the maximum likelihood lens mass and halo fraction,
the dataset is likely to favor even more strongly models with $M = 0$
as the most probable prior, since as the number of events grows more
of the likelihood will concentrate around the maximum likelihood
solution.  Figure~6b shows the evolution of the outcome discriminator
with threshold $\Rcal_*$ for $N = 3$. Clearly, $M = 3$ binary events
is strongly disfavored by the data, even for models in which such an
outcome has a prior relative probability as low as $\Rcal (3|3) \simeq
0.05$. Current data also argue against models in which $M = 2$ has
more than a $30\%$ prior relative probability ($\Rcal >
0.3$). Irrespective of whether current data indicates $M = 1$ or 2, if
the next binary caustic crossing event is inferred to be of non-halo
origin, it will be difficult to account for it with a minimal
Magellanic model. If it is instead found to lie in the Galactic halo,
and one of the two current caustic crossing candidates is also
confirmed to be of halo origin, then the data would be perfectly
consistent with a minimal Magellanic model.

Figures~7a and 7b show the analogous situation for a maximal
Magellanic model (i.e., allowing for the existence of dark Magellanic
halos). The difference between figures~6a and 7a is dramatic. Now
models in which $M=1$ is the most probable prior contain the
overwhelming portion of the likelihood, and models in which $M = 0$ is
expected occupy only the margins of the likelihood. Integrating the
likelihood in each region, we find that we are 55 times more likely to
be living in a galaxy in which $M = 1$ is expected than one in which
$M = 0$ is expected, 95 times more likely to be living in such a
galaxy than one in which $M = 2$ is expected, and 2100
times more likely to be living in such a galaxy than one in which $M =
3$ is favored {\em a priori}\/. Again, as the data accumulate it is
likely to favor models which predict $M = 1$ even more strongly than
suggested by figure~7a, assuming the event timescales observed to date
are representative of the underlying distribution. The outcome
discriminator is shown versus threshold $\Rcal_*$ in figure~7b. As
well as confirming the above findings, the figure shows that models in
which $M = 0$, 1 or 2 have significant prior relative probability
$\Rcal$ are all consistent with the timescales of the current
dataset. Even $M = 3$ is marginally consistent with our adopted cuts,
though it is clearly not strongly favored. If one of the two current
binary candidates is of halo origin, then the discovery of another
non-halo binary would present no problem for a maximal Magellanic
model. Even if the present lenses both reside outside the Galactic
dark halo, a third such event would not represent a major crisis for
the model, although it would be unexpected. If the third binary event
is found to be of halo origin, then there is no problem regardless of
the origin of the first two binary events.

\section{CONCLUSIONS}

The recent discovery of the binary caustic crossing event towards the
SMC has provoked intense debate. It seems certain that this event is
not caused by a halo lens but rather by one in the SMC itself. This
followed the earlier discovery of a binary event towards the LMC,
which also may not lie in the halo. Let us stress that there is some
uncertainty as to the status of the earlier event.  The lens may still
come from the halo, rather than the LMC, if the source star is itself
binary. There is at most a $10 \%$ chance that this is the case. The
alternative is that the lens lies in the LMC. Given the measured
projected velocity $\vproj \approx 20\,\kms$, then this also has a low
prior probability if the velocity dispersion $\sigma$ in the LMC disk
is $\gta 25\,\kms$.  This claim is weakened if either the LMC disk is
colder or thinner than usually assumed.  The ambiguity could be
cleared up by multi-epoch spectroscopy of the source.  Thus, at least
one and maybe both of the events are caused by non-halo lenses.  Note
that our viewpoint differs from that of Sahu \& Sahu (1998), who have
claimed that both the binary and the single events seen towards the
SMC are caused by self-lensing. As evidence for the single event, they
argue that the mass of the lens exceeds 2 $\msun$. Given the absence
of any detectable light from the lens in the spectrum of the source,
there is no easy explanation of what such an object could be.
However, Figure 3 of Afonso et al. (1999) indicates that both the
combined MACHO LMC and EROS SMC data are consistent with masses
between $0.1$ and 1 $\msun$. In this case, it seems premature to
conclude that the lens does not lie in the halo.

In this paper, we have incorporated the valuable information provided
by these exotic events into a statistical analysis of the LMC
dataset. Given the two binary events, we have calculated the prior
probability, as a function of lens mass and halo fraction, that at
least one and possibly both are of non-halo origin for standard
Galactic and Magellanic Cloud models. The lens mass and halo fraction
in these models are themselves constrained by a multi-component
likelihood analysis of the microlensing dataset.  We develop a new
statistic -- the ``outcome discriminator'' -- which allows rigorous
comparison between the likelihood analysis and the prior expectation.
Outcomes in which both binary events lie outside the halo cannot be
made consistent with both prior expectation and the LMC
dataset. Outcomes in which only one of these events is of non-halo
origin provide a reasonable level of consistency.  If both events do
turn out to be of non-halo origin, then one may be forced to consider
alternatives. One possibility is that the Magellanic Clouds themselves
are enveloped by their own dark halos.  Other alternatives are that
the halo may be under-endowed with binaries, or that there is a
serious selection bias favoring non-halo lenses (Honma 1999).  There
is independent support from kinematical studies for the existence of
Magellanic halos. Hitherto, the LMC halo has perhaps not received the
attention it deserves, especially with regard to modeling. Gould
(1993) has shown that an LMC halo could be identified by its
microlensing asymmetry, although this does require more than a hundred
events. We find that the LMC halo makes a significant contribution to
the microlensing optical depth -- around a quarter of the total.
Honma (1999) has argued that the selection effect is severe, with most
halo events missed owing to the typically short time interval between
caustic crossings ($< 10$ days). This effect assuredly exists,
although 10 days is perhaps a little on the long side, allowing for
both bad luck and bad weather after the first caustic crossing. A more
optimistic figure is perhaps 5 days. However, even if the events are
not recognized in real time as binary caustic crossing, they still
would be present as unresolved events in the dataset and could be
searched for (although of course the projected velocity would not be
available).
 
This paper takes the first steps to include the binary caustic
crossing events in the analysis of the microlensing events towards the
LMC.  Given the modest size and the slow growth of the datasets
towards the Magellanic Clouds, it is particularly important to exploit
the second-order information provided by such exotic events.

\acknowledgments
We thank Andy Gould, Geza Gyuk, Prasenjit Saha, Will Sutherland and
HongSheng Zhao for helpful conversations and suggestions.  NWE is
supported by the Royal Society, while EJK acknowledges financial
support from PPARC (grant number GS/1997/00311).

\eject

\begin{table*}
\begin{center}
\begin{tabular}{rccccc}
Component&$\sigma/\kms$& $\tau/10^{-7}$& $\langle t_0 \rangle/$days 
& $\Gamma/10^{-7}\,\mbox{yr}^{-1}$ 
& $N_{\rm exp}(0.35\,\msun)$ \\
\tableline
Galactic halo &$155$&$5.6$ & $65\,(m/\msun)^{1/2}$ &
$20.0\,(\msun/m)^{1/2}$ & $16$ \\
Galactic disk &$30$ & $0.06$ & $92\,(m/\msun)^{1/2}$ &
$0.14\,(\msun/m)^{1/2}$ & $0.12$ \\
LMC halo &$55$&$2.1$  &$95\,(m/\msun)^{1/2}$ &$5.1\,(\msun/m)^{1/2}$ 
& $4.1$ \\
LMC disk &$25$ &$0.54$ & $87\,(m/\msun)^{1/2}$ &
$1.4\,(\msun/m)^{1/2}$
& $1.2$ \\
\null &\null&\null&\null&\null&\null\\
\end{tabular}
\end{center}


\tablenum{1}
\caption{This table gives the velocity dispersion $\sigma$, the 
optical depth $\tau$, the average timescale (Einstein radius crossing
time) $\langle t_0 \rangle$ and the rate of microlensing $\Gamma$
towards the Large Magellanic Cloud for a number of possible lensing
populations. The expected number of events $N_{\rm exp}$ for the
2-year Alcock et al. (1997b) dataset is also given for a lens mass of
$0.35\,\msun$, incorporating their efficiencies.}

\end{table*}

\eject

\begin{figure}
\begin{center}
               \epsfxsize 0.6\hsize
               \leavevmode\epsffile{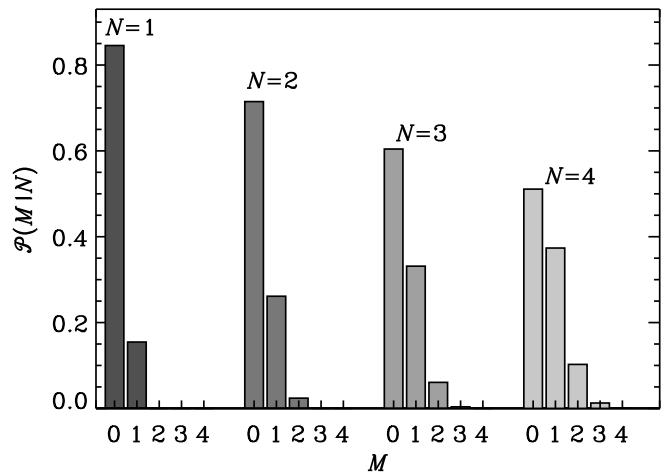}
\caption{The prior probability of observing $M$ non-halo caustic
crossing binaries from a total sample of $N$ such events. The plot
assumes a binary bias $B=1$ and a rate factor $R=5.5$. Current
observations are best described by $N=2$ and either $M = 1$ or 2.}
\end{center}
\end{figure}

\eject

\begin{figure}
\begin{center}
               \epsfxsize 0.75\hsize
               \leavevmode\epsffile{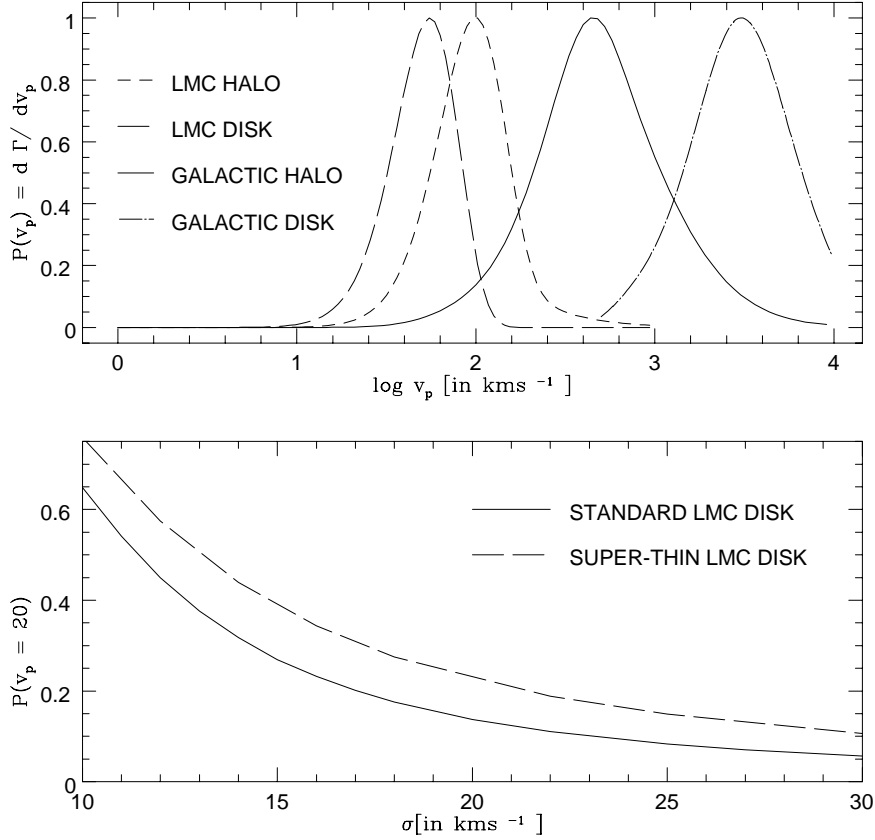}
\caption{
The upper panel shows the distribution of projected velocities
$P(\vproj)$ for lenses lying in the Milky Way halo and disk, as well
as the LMC halo and disk. Each of the distributions has been
normalized so that its maximum lies at unity. As Bennett et al. (1996)
already noted, this diagram gives a clean separation between the
locations of the lenses. The lower panel assesses the probability of
finding a lens with projected velocity as low as $\sim 20\,\kms$ as a
function of the velocity dispersion of the LMC. The unbroken line
refers to a model of the LMC with the standard scale height of
$300\,\pc$ while the broken line is for a super-thin LMC disk with
half the standard scale height. The vertical axis is the probability
density $P(\vproj = 20)$ measured relative to the most common
projected velocity.}

\end{center}
\end{figure}

\eject

\begin{figure}
\hspace*{3.5cm}
\rotate[r]{
               \epsfxsize 0.45\hsize
               \leavevmode\epsffile{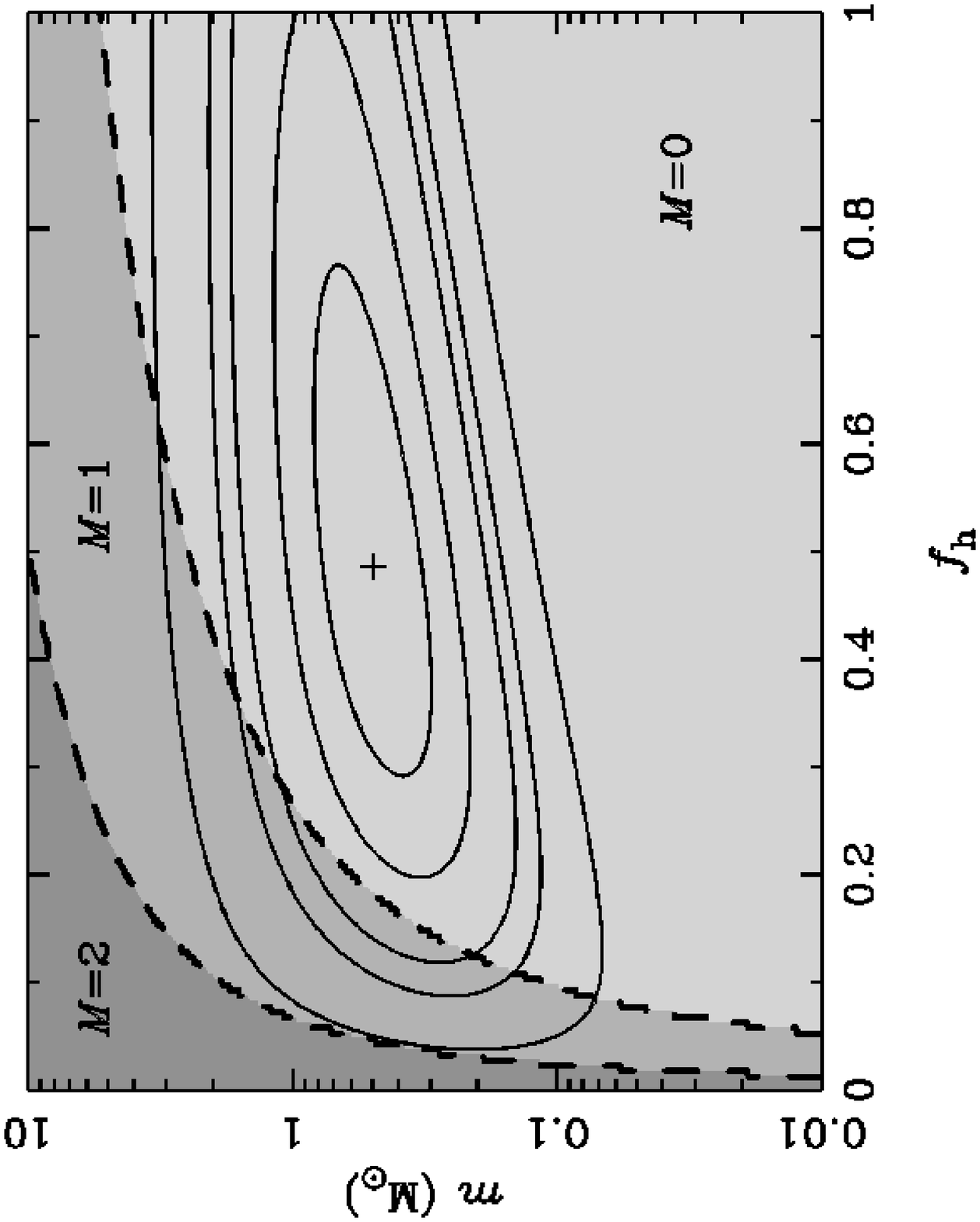}
               \epsfxsize 0.4\hsize
               \leavevmode\epsffile{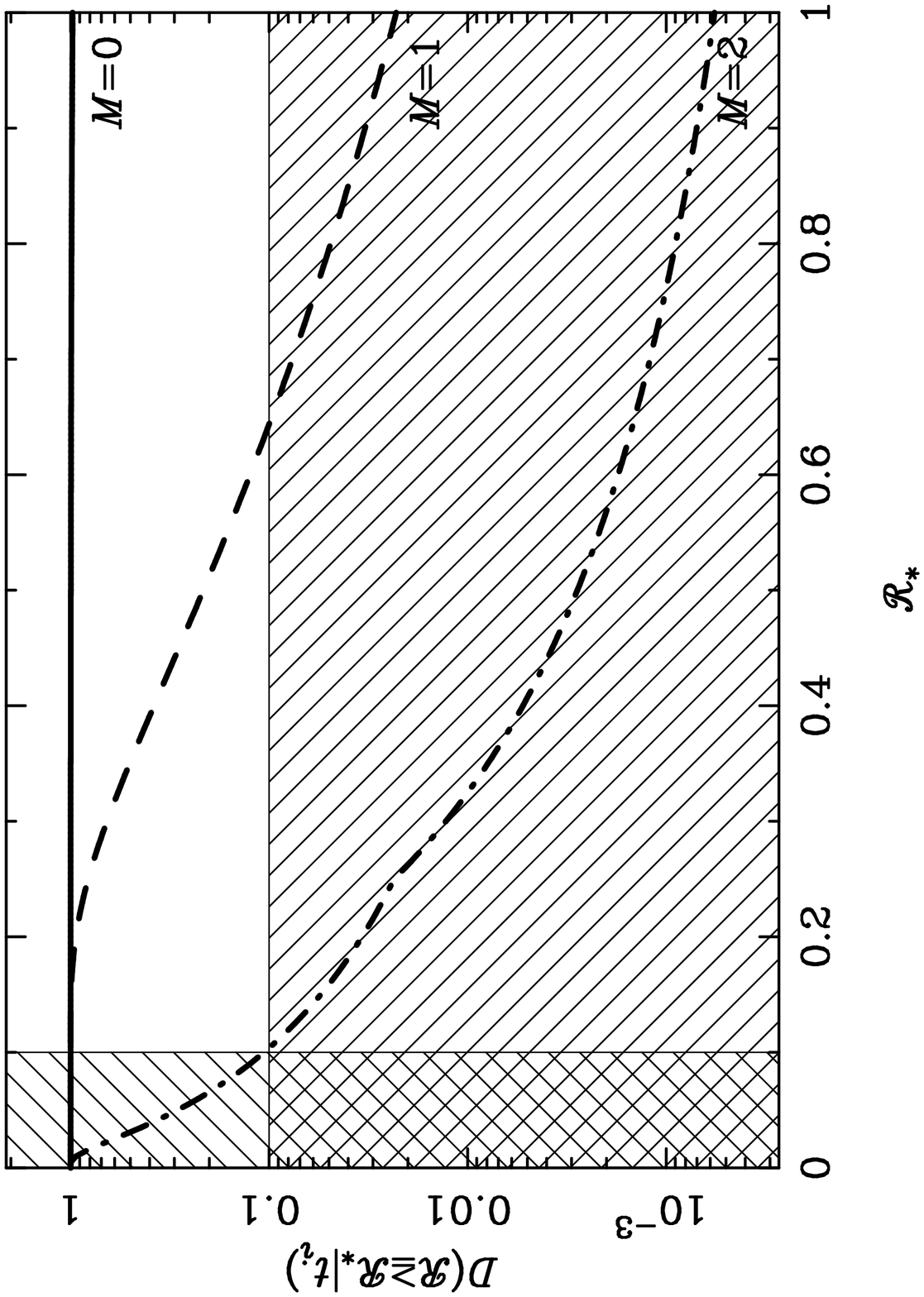}
}
\caption{(a) The solid lines show the $34 \%$ (innermost), $68 \%$,
$90 \%$, $95 \%$ and $99 \%$ (outermost) likelihood contours in the
plane of lens mass $m$ and halo fraction $\fha$ resulting from the
Alcock et al. (1997b) 8 event LMC dataset, assuming a combined
Galactic (disk+halo) and minimal LMC (disk-only) model. The shaded
regions indicate models in which $M = 0$, 1 or 2 non-halo binary
caustic events are expected from $N = 2$ observed caustic crossing
binaries. (b) The ``outcome discriminator'' $D(\Rcal \ge \Rcal_* |
t_i)$ (see main text) as a function of prior relative probability
threshold $\Rcal_*$ for the case $N = 2$ and $M = 0$ (solid line), 1
(dashed line) and 2 (dot-dashed line). The hatched, vertical and horizontal
areas respectively denote regions of unacceptably low prior
relative probability given the model ($\Rcal < 0.1$) and unacceptably low
likelihood given the two-year LMC data [$D(\Rcal \ge \Rcal_* |
t_i) < 0.1$].}
\end{figure}

\eject

\begin{figure}
\hspace*{3.5cm}
\rotate[r]{
               \epsfxsize 0.45\hsize
               \leavevmode\epsffile{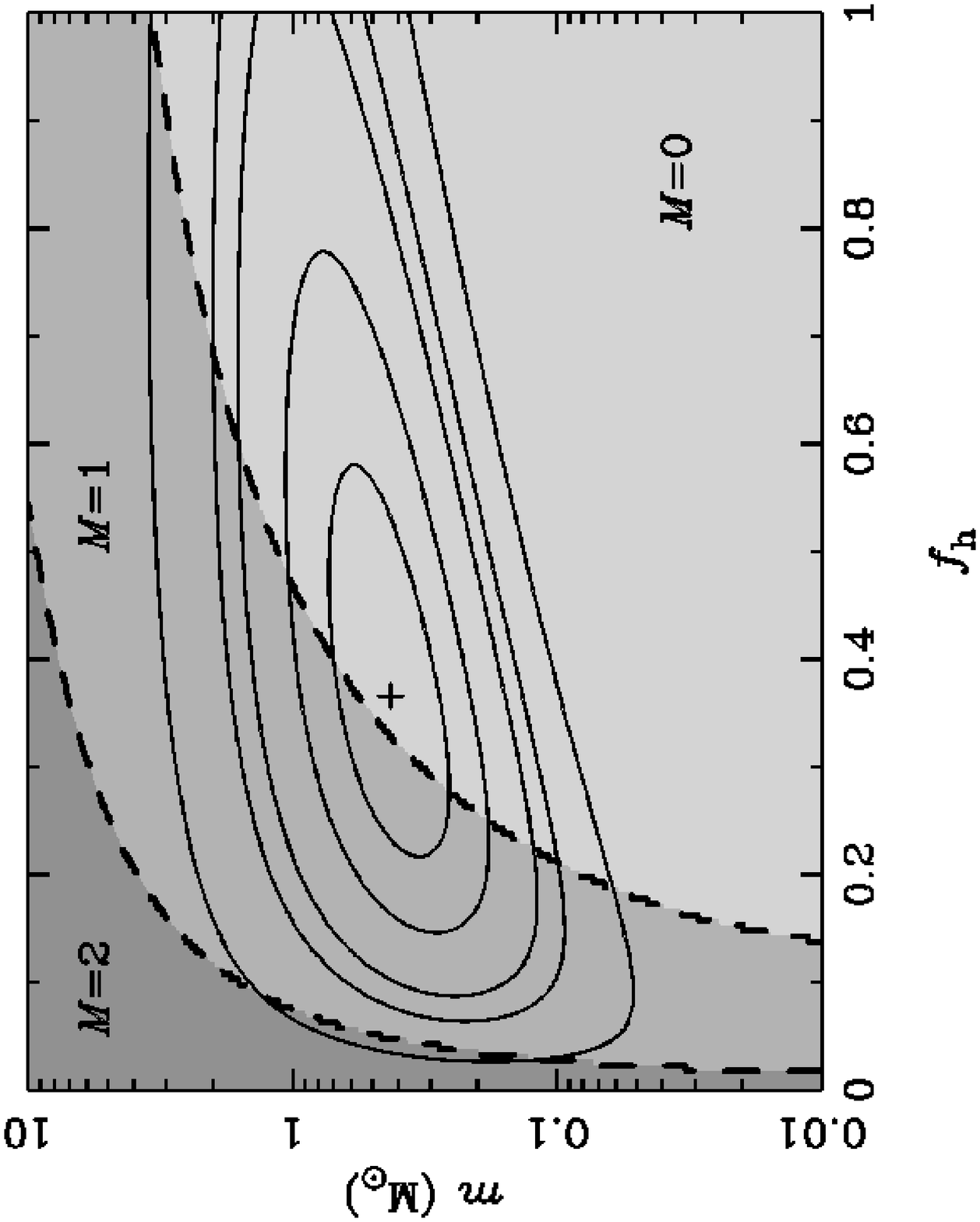}
               \epsfxsize 0.4\hsize
               \leavevmode\epsffile{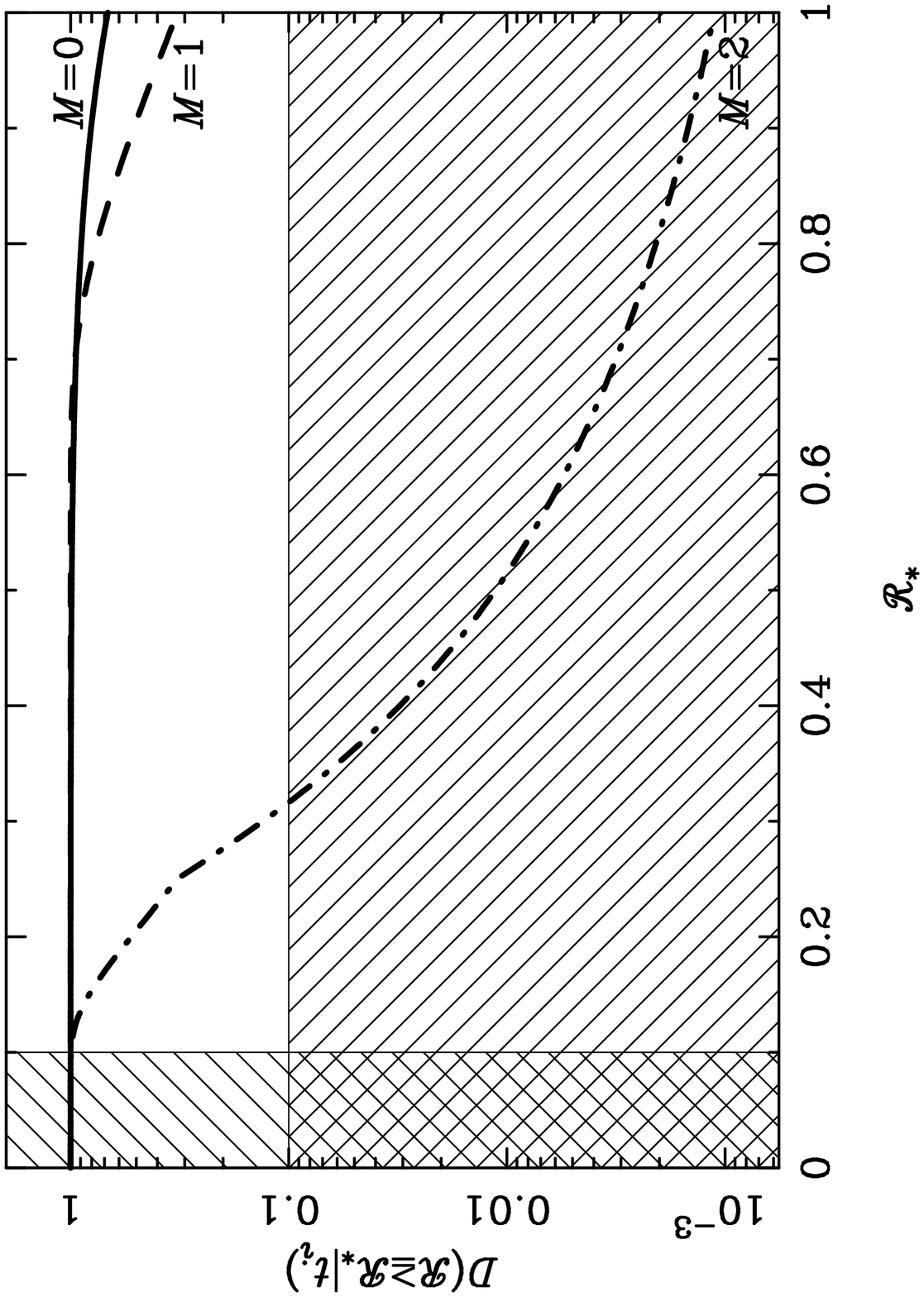}
}
\caption{As for figure~3, but for a maximal model in which the LMC
possesses a dark halo of its own, characterized by the same baryon
fraction $\fha$ and lens mass $m$ as the Milky Way halo. Notice that
the ``M=1'' region now occupies a more substantial share of the total
likelihood.}
\end{figure}

\eject
\begin{figure}
\hspace*{3.4cm}
\rotate[r]{
               \epsfxsize 0.45\hsize
               \leavevmode\epsffile{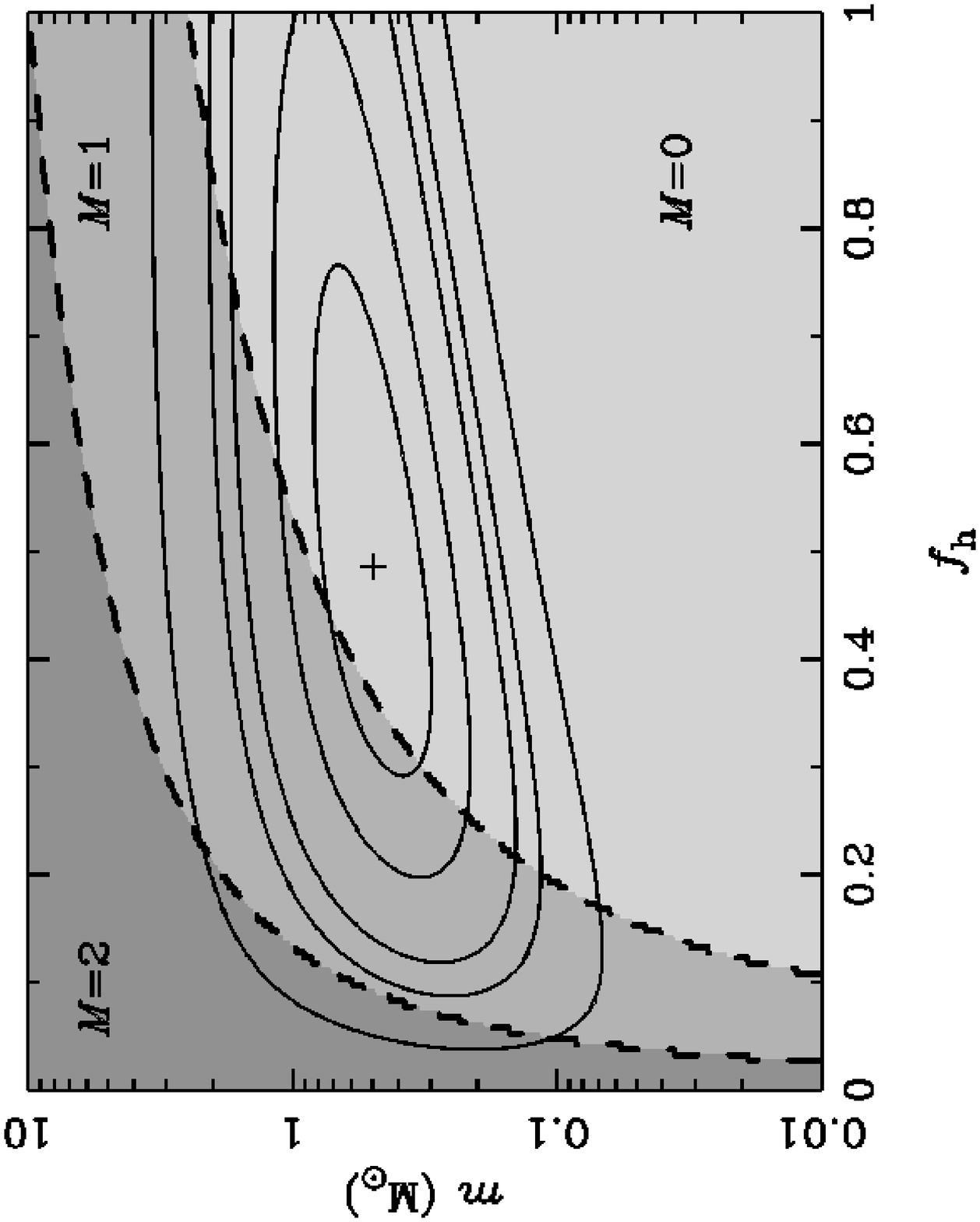}
               \epsfxsize 0.4\hsize
               \leavevmode\epsffile{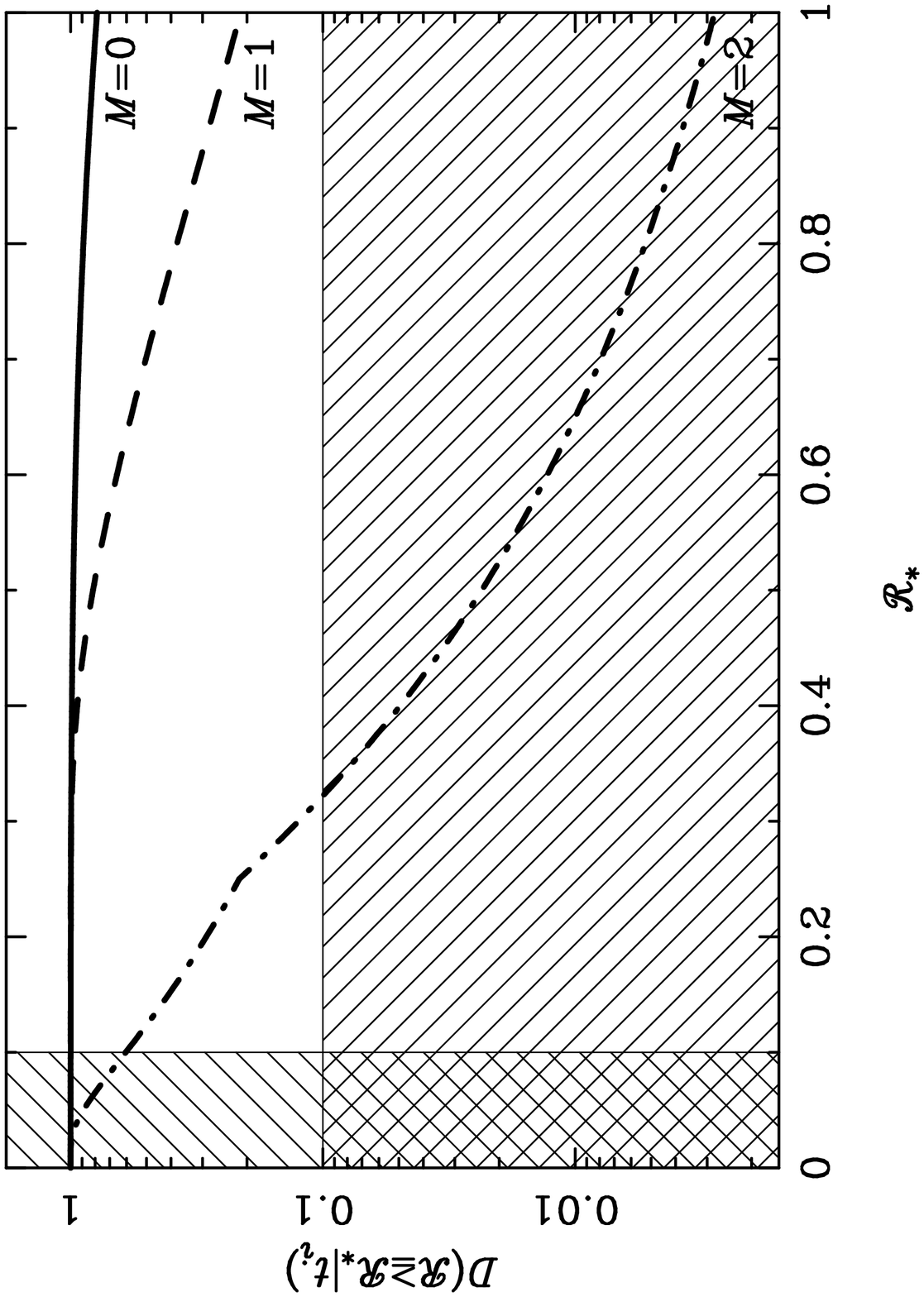}
}
\caption{As for figure 3, but with a binary bias $B = \half$. This
means that the caustic crossing binary fraction is twice as large
for the LMC and Milky Way disk than for the Milky Way halo. Again, the
``M=1'' region shares more of the likelihood than in figure 3.}
\end{figure}

\eject

\begin{figure}
\hspace*{3.5cm}
\rotate[r]{
               \epsfxsize 0.45\hsize
               \leavevmode\epsffile{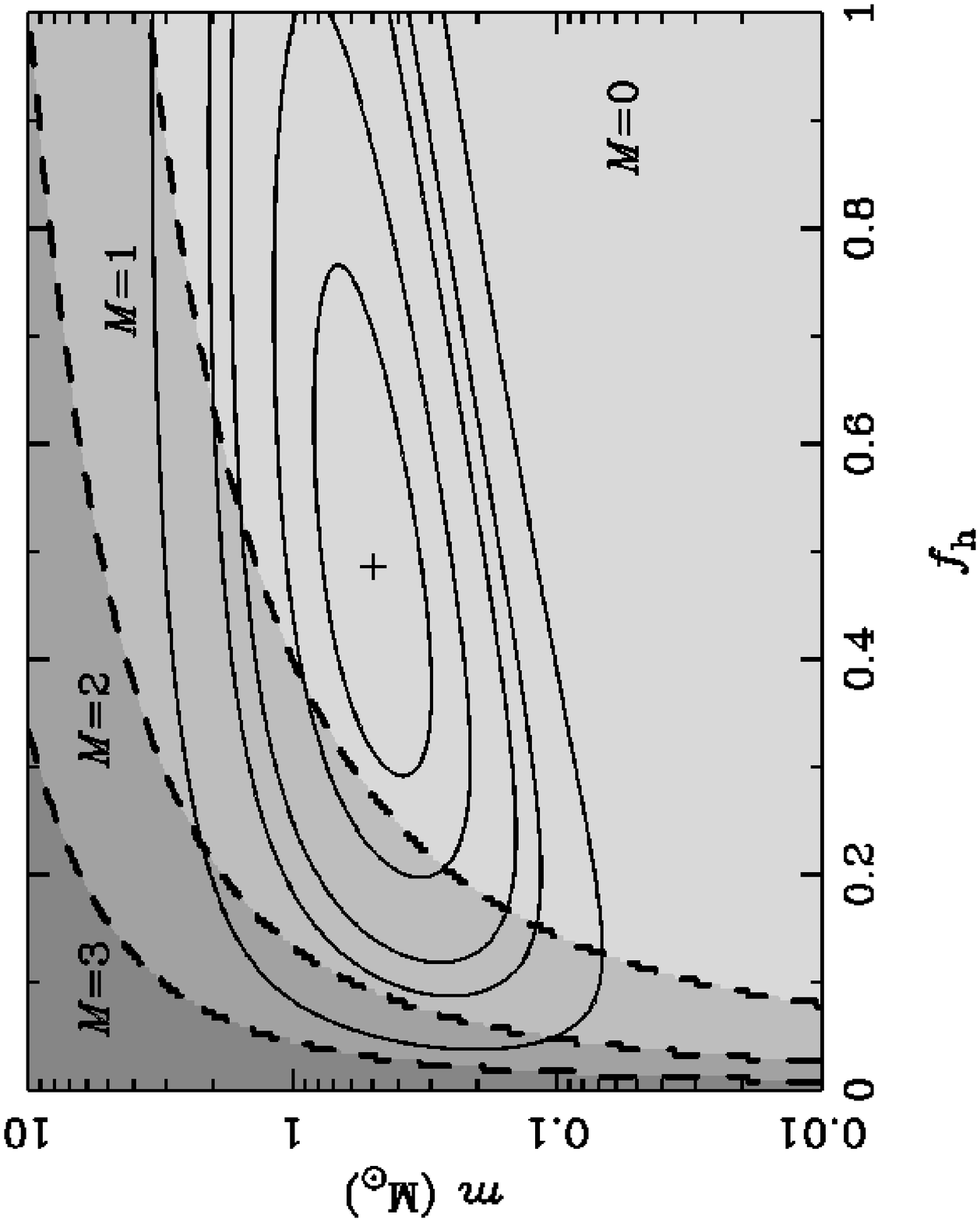}
               \epsfxsize 0.4\hsize
               \leavevmode\epsffile{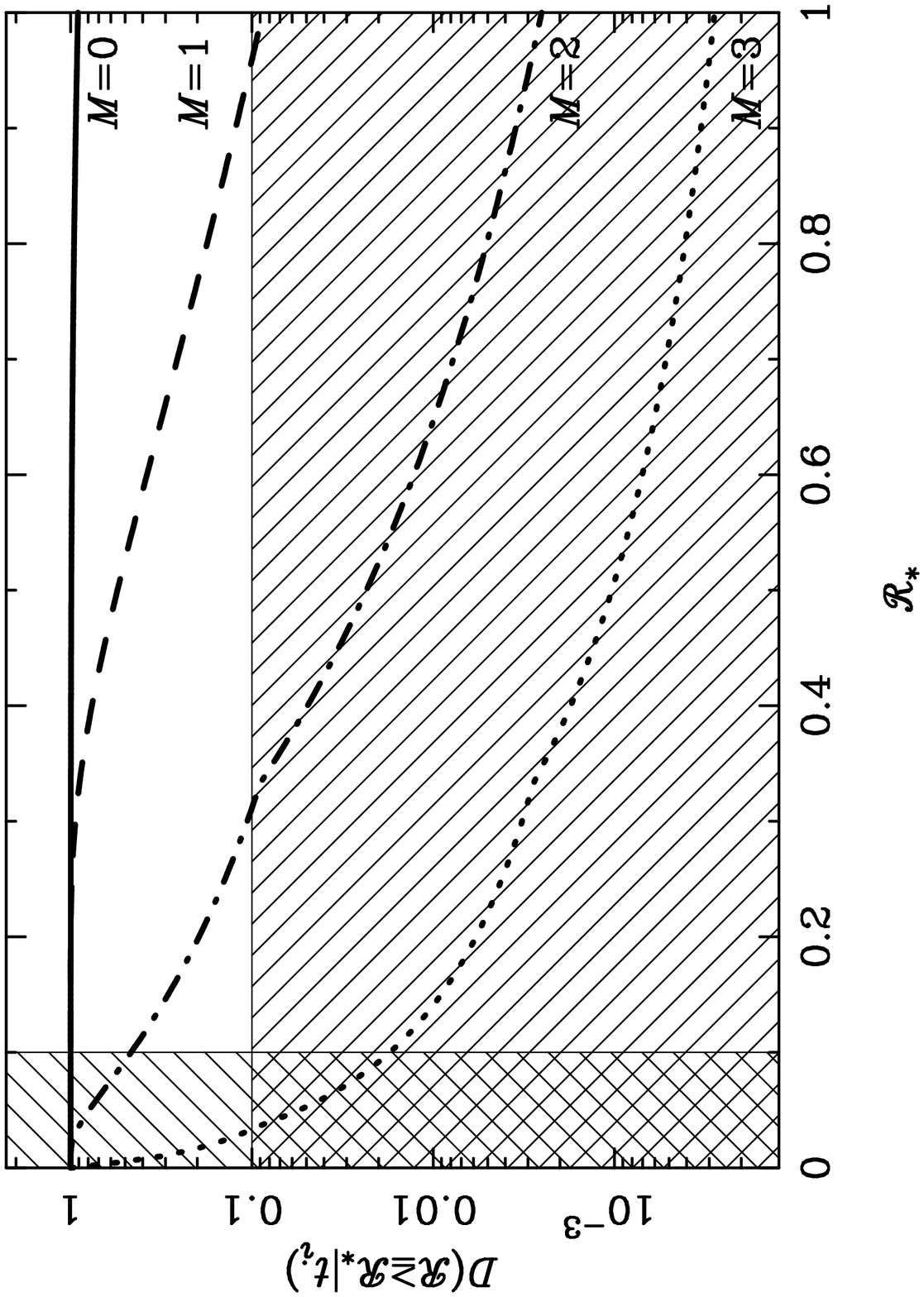}
}
\caption{As for figure 3, but for the instance $N = 3$. So, this 
figure assumes that a further binary caustic crossing event has been
discovered.  The LMC is modeled by a bare disk. Notice that $M=0$ is
still the most likely outcome.  Irrespective of whether current data
indicates $M = 1$ or 2, if the next binary caustic crossing event is
inferred to be of non-halo origin, it causes difficulties for this
model. }
\end{figure}

\eject

\begin{figure}
\hspace*{3.4cm}
\rotate[r]{
               \epsfxsize 0.45\hsize
               \leavevmode\epsffile{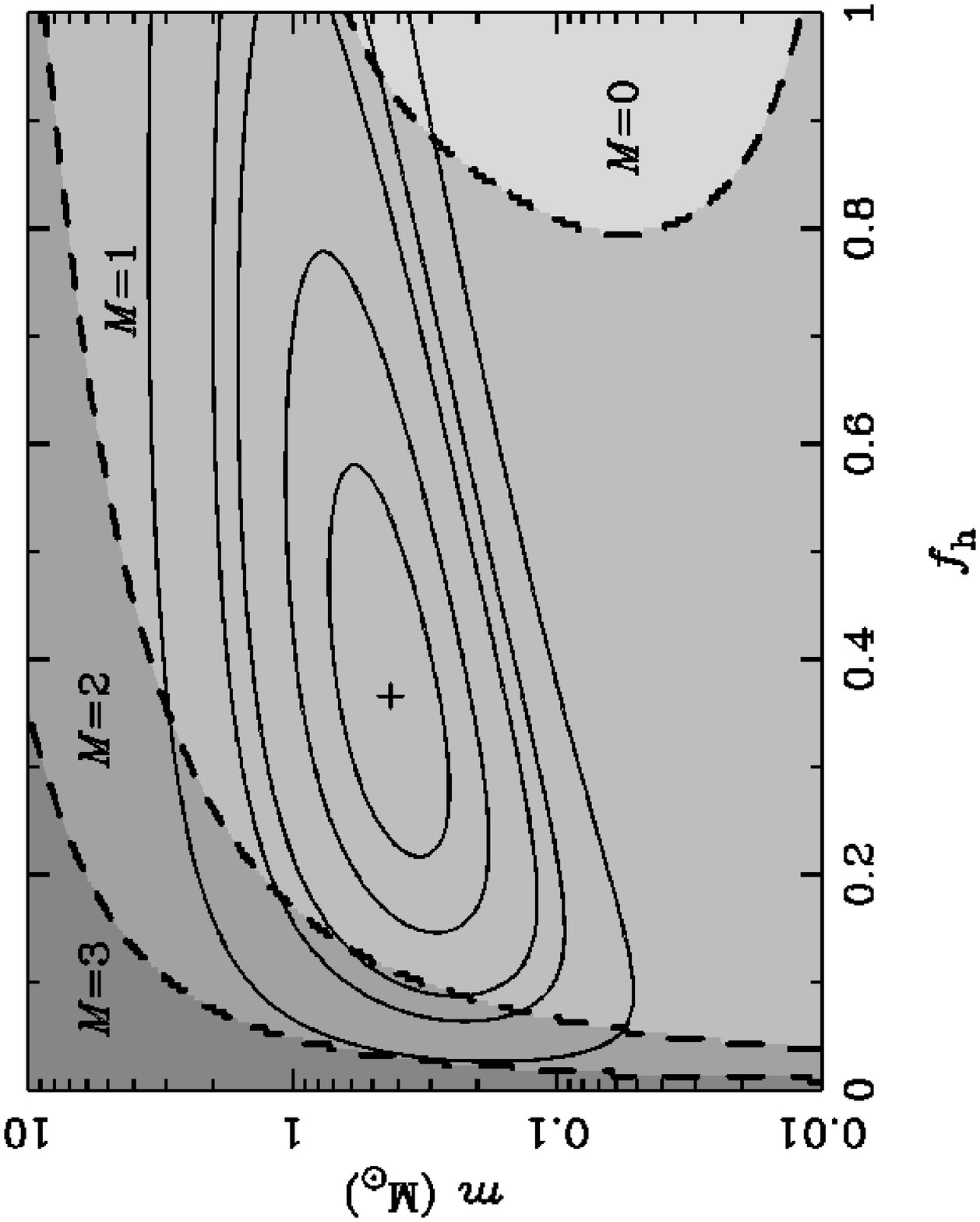}
               \epsfxsize 0.4\hsize
               \leavevmode\epsffile{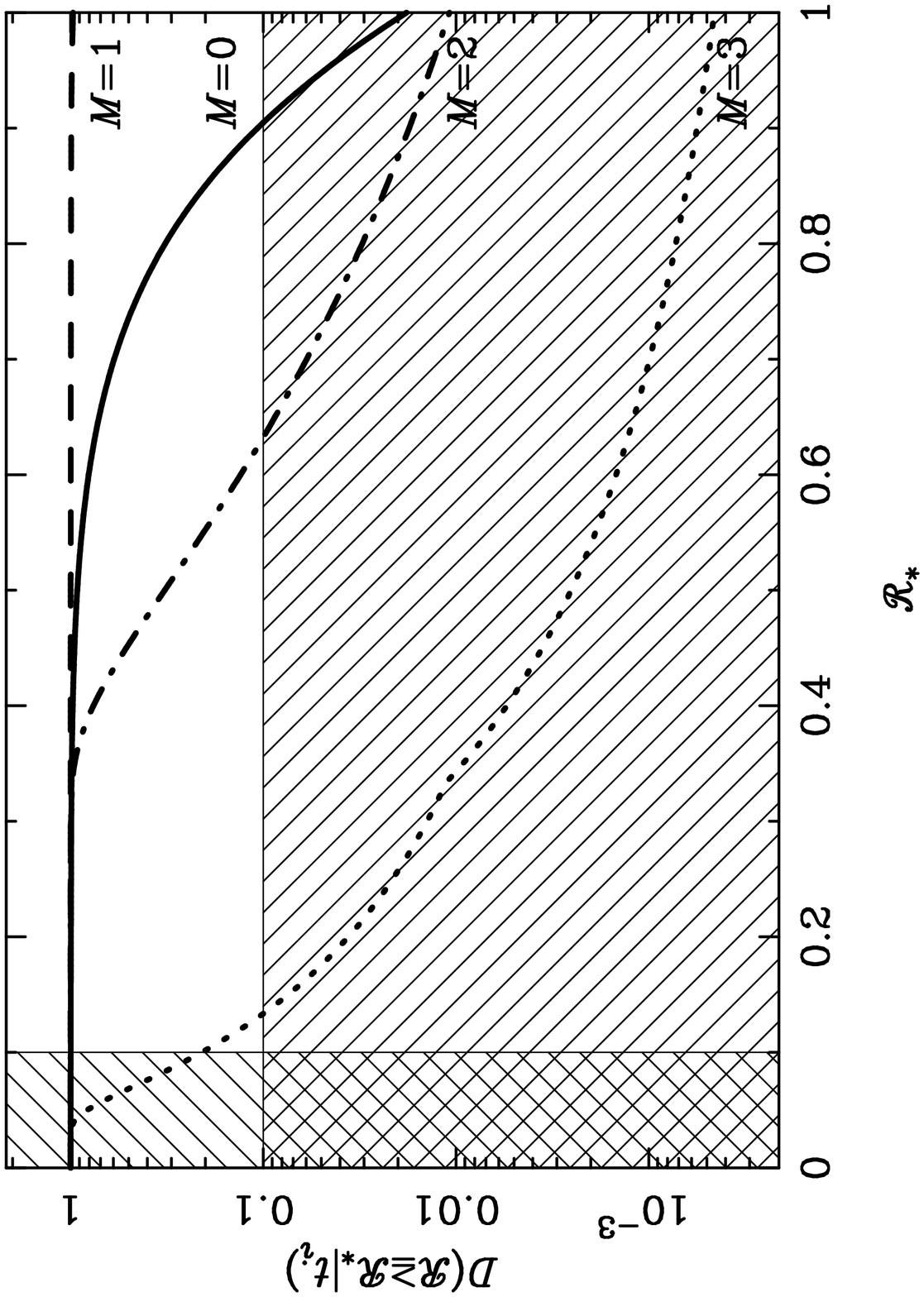}
}
\caption{As for figure 6, but using a model for the Magellanic Clouds
that includes an enveloping dark halo. The most likely outcome is
now $M=1$. Even if a further event is discovered, there is no serious
inconsistency whether or not it is found to be caused by a halo or
LMC lens.}
\end{figure}

\eject

\end{document}